\documentstyle[twocolumn,aps]{revtex}
\begin{document}
\title{
\begin{flushright}
\small
\end{flushright}
A study of supercooling of the disordered vortex phase via minor
hysteresis loops in   2H-NbSe$_2$\\
\vskip 0.25truecm
\normalsize
\noindent
G. Ravikumar$^{1,*}$, P. K. Mishra$^1$, V. C. Sahni$^1$, S. S.
Banerjee$^2$, A. K.
Grover$^2$, S. Ramakrishnan$^2$, P. L. Gammel$^3$, D. J. Bishop$^3$,
E. Bucher$^3$, M. J. Higgins$^4$ and S. Bhattacharya$^{2,4,*}$
\vskip 0.1truecm
{\it
$^1$TPPED, Bhabha Atomic Research Centre,
Mumbai-400085, India\\
$^2$ Dept. of Condensed Matter Physics and Materials Science, Tata
Institute of Fundamental Research,
Mumbai-400005,
India\\
$^3$Bell Laboratories, Lucent Technologies, Murray Hill, NJ 07974\\
$^4$ NEC Research Institute, 4 Independence
Way, Princeton, NJ 08540\\}
\vskip 0.2 truecm
\small \noindent
We  report  on the observation of novel features in the minor hysteresis
loops in a clean crystal  of  $NbSe_2$  which
displays a peak  effect.  The observed behavior can be
explained in terms of a supercooling of the disordered vortex phase while
cooling the superconductor in a field. Also, the extent of spatial order
in a flux line lattice formed in  ascending  fields  is different from (and
larger  than) that in the descending fields below the
peak position of the peak effect; this is attributed to unequal degree of
annealing of the state induced by a change of field in the two cases.
\small
\begin{center}
PACS numbers :64.70 Dv, 74.25 Ha 74.60 Ge, 74.60 Jg
\end{center}
}
\maketitle
\normalsize
\bigskip
\vskip 0pt
\normalsize
\noindent
\section{INTRODUCTION}
A variety of recent transport, magnetic and structural studies in weakly
pinned superconductor
2H-NbSe$_2$ support the view that the peak effect (PE) phenomenon  [1]
in the critical current density ($J_c$) marks an amorphisation of the
flux line lattice (FLL)  [2-10]. In a varying field experiment, for example,
[6,9],  $J_c(H)$  begins to increase from an initial low value at  the
onset field $H_{pl}^+$, reaches a
maximum at the peak field $H_p$ and  eventually collapses below the
measurable limit above the
irreversibility field  $H_{irr}$  (  $<$  $H_{c2}$). The vortex matter
is thought to undergo a
transformation from a state with a high spatial order for  $H$  $<$
$H_{pl}^+$  to a highly disordered state for $H$ $>$ $H_p$. This
interpretation is usually rationalized
within  the  Larkin-Ovchinnikov  (LO) collective  pinning
formalism [11], where the Larkin volume $V_c$ ($\approx$ $R_c^2$ $L_c$, where
$R_c$ and
$L_c$  are  the  radial  and  longitudinal  correlation  lengths,
respectively)  is  a measure of the spatial extent of order in
FLL and the critical current density is determined through the  relation,
$\mu _0 H J_c(H)$  $\approx$ $(n_p <f^2> / V_c)^{1/2}$, where $n_p$ and
$f$ are pin
density and the elementary pinning force parameter,
respectively.  $J_c$ measurements therefore can reveal the extent of
spatial order in the FLL. Magnetization
hysteresis measurements are  a convenient  tool for estimating
$J_c(H)$ and thus for detecting the occurrence of phase transformations
and associated changes in the vortex correlations across the PE regime.

Recently, the PE phenomenon has received a great deal of attention due to
a characteristically rich phenomenology that accompanies it. Prominent
among them is the marked history dependence in the critical current. In
the LO scenario, this translates into a history dependence of the Larkin
volume $V_c$, since quantities such as $f$ and $n_p$ cannot be history dependent.
The history dependence of the correlations implies strong metastability
and is thus a hallmark of disorder in condensed matter systems. Recent
theoretical studies emphasizing the role of quenched random disorder in
producing novel disordered (glassy) phases (see, for instance, Refs. 12-14) 
further illustrate the need to
understand and ultimately unravel the complex effects of disorder readily
observed in the PE regime.
Ravikumar et al [9] have experimentally demonstrated via dc magnetization technique the presence of a
highly disordered vortex state when a sample is field cooled (FC) in $H$
$<$ $H_p$[6]. In crystals of
$2H$-$NbSe_2$ and $CeRu_2$ with comparable levels of effective pinning,
they had  shown  that the
critical current density in the FC state is larger than that in the zero
field cooled (ZFC) state for  $H$
$<$  $H_p$ , i.e., $J_c^{FC}(H)$ $>$ $J_c^{ZFC}(H)$. They had also shown
that the disordered
FC state could be annealed to the ordered ZFC like state when the sample
was subjected  to  a  small  change  in magnetic field.
Independently, a different type of history effects has been reported [15,16] in
a wide
variety of polycrystalline and single crystal samples of pure and
doped $CeRu_2$ showing PE phenomenon. In these cases, the minor magnetization curves, starting at a field $H$ on
the forward branch of the hysteresis loop
(such that $H_{pl}^+$ $<$ $H$ $<$ $H_p$), were reported to saturate
without merging
with the reverse leg of the hysteresis loop. The minor curves
starting on the
reverse branch, on the other hand, merge with the forward branch of the
hysteresis loop. Roy et  al  [15]  attribute
the observed behavior to an {\it anomalous} nature of
PE phenomenon in the mixed valent superconducting system $CeRu_2$, in
contrast with, and as distinct from,
the conventional PE in  $2H$-$NbSe_2$[2] and most other  weakly
pinned low $T_c$ superconductors [9,10,17]. They propose [15,16] that this novel
behavior of the minor loops reflects a thermodynamic evidence for a first order transformation to a new phase caused by the positional dependence of the order parameter in superconductors with large normal state paramagnetism, like, heavy fermion superconductors, mixed valent rare earth systems, etc. [18]. This is in contrast to the explanation based on metastability effects in the vortex matter caused by quenched disorder.

Very recently, a similar effect with the minor hysteresis curves has been
observed in a cuprate superconductor as well [19]. The microscopics in the cuprate system appears to bear little resemblance to that in the mixed valent systems such as CeRu$_2$.
We propose that the above sets of experiments  [9,15,16,19]  exemplify the
ubiquitous nature of thermomagnetic history dependence of $J_c$ in
superconductors in general and thus require an explanation that is
independent of the microscopics relevant for superconductivity in the
diverse systems in which these effects are present. 

In this paper, through detailed
measurements of minor magnetisation curves on a clean single crystal
of $2H$-$NbSe_2$ (belonging to the category  of  most  weakly
pinned   samples  of  type  II  superconductors [2,4]), we  present an understanding,
based on the LO collective pinning description
[11] for the observed path dependence in the critical currents and vortex correlations. We
invoke the notion of a disorder-assisted
supercooling of a metastable disordered phase, that is otherwise
thermodynamically stable only above $H_p$. We further propose that changes
in the applied field act as a driving force that helps to anneal the
system, often partially. As a result, systems with different field
histories are often in metastable states with different
degrees of annealing and thus with different values of vortex
correlations, leading to the different critical currents, as observed
experimentally.
In the present work, three types of isothermal minor hysteresis loops
were studied : (I) Decreasing field after cooling the sample in a field (FC-REV),
(II) and (III) : decreasing /increasing field from a given point on the
forward/reverse leg of the envelope hysteresis curve. Fig.1 shows a
schematic view of all these loops for the case where $J_c(H,T)$ is
single-valued, independent of the field/temperature history. In all such
cases, the minor curves merge into the envelope curve while always
remaining within it, without any overshoot effects. We show below that
violations of this standard scenario provide an understanding of the
aforementioned anomalous behavior.

\section{EXPERIMENTAL}
We  carried out {\it dc} magnetization measurements using a Quantum
Design (QD) SQUID magnetometer (Model MPMS5) on a $2H$-$NbSe_2$ single  crystal
with $T_c$ $\approx$ 7.25 K. The crystal was  mounted  with  field
applied  parallel  to  its {\it  c}-axis.  The thermomagnetic history
dependent measurements have been performed at 6.95 K, where a  well
recognized  peak  effect,  manifesting as a sharp increase in the
magnetization hysteresis, is observed at a relatively  low  field
of about 1000 Oe. Occurrence of PE at such low fields, where
the flux line lattice constant $a_0$ (= 1600 A) is  of  the  same
order as  the  range  of  interaction  ( i.e.,  the penetration depth
$\lambda _{ab}$) at 6.95K in this system [20], confirms that the
sample has weak quenched disorder [7]. In the field
range 1$-$2 kOe, the inhomogeneity experienced by the sample in a 2 cm
full-scan  in
a QD  SQUID  magnetometer  is of the order of 0.1 Oe[21]. This value is
much smaller than the threshold field $H_{II}$ [22]  required  to
change  the sign of the induced shielding currents throughout the
sample. $H_{II}$ value is estimated  by  measuring  the
minor  magnetisation curves [23,24] for fields above $H_p$ and was found to be about 10 Oe. Thus,
the field
inhomogeneity in a full scan  of 2  cm  does not introduce  any
error in  the  measured  magnetization  values,  and  we  have
recorded all the present data with a 2 cm scan  length  instead  of
{\it half  scan  technique} utilized earlier by Ravikumar et
al [9,21], in their study of $CeRu_2$ and  more  strongly  pinned
sample  of 2H-NbSe$_2$.

\section{RESULTS AND DISCUSSION}
In   Fig.2,   we   show  the  magnetization  hysteresis  loop in the present 2H-NbSe$_2$ crystal at 6.95
K, indicating the onset field of PE  on the ascending
field  cycle ($H_{pl}^+$ $\approx$ 800 Oe), the peak
field $H_p$ ($\approx$ 1000 Oe) and the irreversibility field  $H_{irr}$
($\approx$ 1250 Oe). Within the LO collective pinning description,
the Larkin volume  $V_c$  begins  to  shrink  at  $H_{pl}^+$  and FLL
reaches nearly amorphous (though
pinned) state at $H_p$ [7,10]. Note that  the
onset of PE is much sharper (see inset(ii)) on the ascending field cycle than
on  the  descending field cycle. This immediately shows that at a
given field value in the PE region, $J_c$ and thus $V_c$
in the vortex state are not the same on
the ascending and
descending field cycles in the PE regime. In what follows, we examine this
non-uniqueness of the vortex correlations in greater detail.

In Fig.3(a), we show the magnetization curves measured for FC-REV
case,  i.e.,  magnetization  of a field cooled sample measured in
decreasing magnetic fields. The field value in which the sample
was   cooled   was  varied  across  the  PE  region. Note that in
Fig.3(a), the FC-REV curve originating at $H$ $>$ $H_p$
merges  with the  reverse magnetization envelope curve [23,24] in
accordance with the critical  state  model. On
the other hand, FC-REV curves, originating from a field $H$ $<$ $H_p$,
overshoot the reverse magnetization envelope
curve in a clear departure from the conventional behavior
shown in Fig.1. The difference between the highest magnetisation
value recorded on
the FC-REV curve and the notional equilibrium magnetisation value
could be taken as a measure of the critical current density in the FC
state [25]. Thus, the minor curves in the FC-REV case produce higher
$J_c$ values than those in the conventional descending part of the envelope curve.

This anomalous behavior can be explained by assuming a
supercooling of the disordered phase and the annealing effect (i.e., increase in vortex lattice correlations or growth of Larkin domain volume $V_c$) due to a subsequent field
change. In the FC state, the  FLL
traverses  through  a pinned
amorphous  state  as  it  is  cooled  down across  the $H_p(T)$ line
(see the phase diagram drawn in the inset (i) of Fig.2). The highly disordered vortex state that is stable above PE curve,
with a  large
density of defects/dislocations [2,5,7], is then effectively supercooled
when the sample
is cooled to a given T in a field less than H$_p$(T). Vortex state obtained by cooling the sample in a field H
is more disordered than that at the same field value on the descending branch  of the envelope loop. In the latter case, the process of lowering of field (below $H_p$) induces partial
annealing and produces a vortex state, which is more correlated and thus has a
smaller critical current than that in the field cooled state. Furthermore, annealing induced by the field change as mentioned
above is also clearly seen in the FC-REV  magnetization  values moving towards the reverse envelope curve (see inset of Fig.3(a)). 

The  FC-REV curves initiated from different fields below $H_p$
form a family of curves. FC-REV curve originating from $H$ $=$ $H_p$
essentially retraces the reverse magnetisation envelope curve below
$H_p$. As stated earlier, the vortex state at any field value below $H_p$
on  the reverse  magnetization curve is
the result of a gradual healing  of the
disordered state existing at $H = H_p$. Thus, the reverse magnetization
envelope curve (for $H$ $<$ $H_p$), in principle need not, and in
practice will not, be a mirror
reflection of the forward  magnetization envelope curve as the specific
kinetics of the annealing processes are different. That the process of
healing of FLL dislocations could continue down to a field well below
$H_{pl}^+$ (during the descending cycle) is a clear indicator of this difference.
The data shows that for a  given  field  $H  (< H_p)$, the Larkin domains
on the reverse magnetization envelope curve are smaller than those
on the forward magnetization envelope curve. In other words,
$J_c$  values  on  the  descending  field
cycle are larger than those on the ascending field cycle. This difference is likely to be 
due to the difference in the starting configurations on the ascending and
descending branches. For the latter, one starts from a much more
disordered state; thus the residual disorder after a comparable level of
(incomplete) annealing is more than that in the former.

Using the scenario described above, we now examine the minor magnetisation curves of the type II and III (cf. Fig.1) in the PE region of 
2H$-$NbSe$_2$   and   compare  and  contrast  them  with  anomalous
behavior of minor magnetisation curves in the PE region of $CeRu_2$
[15,16]. In Fig.3(b), we first show
the  minor  magnetization  curves   of type II, initiating from
different fields lying
on the forward magnetization curve. The minor magnetization curves
initiated from $H > H_p$ and $H < H_{pl}^-$ merge with the reverse magnetization
curve within a field change of about 10 Oe. The threshold field $H_{II}$ is
thus estimated from these curves to be of the order of 10 Oe.
However, for $H_{pl}^- < H < H_p$, the minor magnetization  curves  do
not  merge into the reverse envelope curve, because $J_c(H)$ values
on the ascending field cycle are smaller than those on  the  descending
field  cycle, as  asserted earlier. The  significance of $H_{pl}^-$ is
that the disorder present at $H = H_p$ is maximally annealed at  this
field value on the descending cycle.

In Fig. 3(c), we show the measured  minor magnetization curves of the type III, i.e., by
increasing  the  field  from different points on the reverse magnetization
envelope curve.  These minor curves now overshoot the forward envelope
magnetisation curve when the field is increased by about 10 Oe. The 
annealing due to the field increase (of about 10 Oe) is inadequate to produce a
comparable level of lattice order existing at the corresponding fields on the forward envelope loop. When the field is further increased, the residual disorder gets annealed. Thus the minor magnetisation curves eventually merge into the more ordered forward envelope curve. However,
Roy et al[15] reported that in $CeRu_2$ the minor curves initiated from
the reverse magnetisation envelope curve readily merge with the forward
magnetisation curve, in contrast with our data. Whether this is a
significant difference, or merely a trivial one, in the comparative
levels of annealing in the two instances, is unclear [26].

The critical current densities, $J_c^{for}$
and $J_c^{rev}$ on the ascending and descending field branches
can now be estimated from the
maximum width of these two sets of minor magnetisation curves. We
collate, in Fig.4, the relative $J_c(H)$ values
corresponding to three thermomagnetic  histories  of  the  sample,
viz., the FC state and the states along the {\it forward} and the
{\it reverse}  legs  of the envelope hysteresis loop. The three sets of
$J_c(H)$ values have been estimated [9,15,25] by  taking  the  notional
half  width  of the magnetization hysteresis at a given $H$ to be
proportional to the corresponding critical current  density.  The
data in Fig.4 can be summarized by the inequality,
$$
J_c^{FC}(H) > J_c^{rev}(H) > J_c^{for}(H), 
$$
which  is consistent with an early report of transport data in  a  Nb
crystal  by  Steingart  et al [27]. In the framework mentioned above,
this inequality corresponds to the least annealed state in the FC mode and
the most annealed state
on the forward curve, while the $J_c^{rev}$ is intermediate between the two.

\section{Conclusion}
We propose that the PE phenomenon marks a true thermodynamic phase
transformation between an ordered solid that is stable below $H_{pl}^+$ (on forward envelope loop) and a fully 
disordered vortex state that is stable above $H_p$. Further, this
transformation is first order in character. Thus, it is possible to
{\it supercool} or {\it superheat} one phase into the regime of stability of the
other phase. For the PE regime, when the free energies of the ordered and
disordered phases are not significantly different, the metastability
effects are expected to be prominent. Moreover, thermal fluctuations are
inadequate, at least below $H_p$, for the system to fully explore the phase
space, which helps the metastability of the {\it supercooled} phase, aided by
the fact that both phases have finite pinning. Substantial driving
forces experienced when the field is changed allow the system
to anneal (or fracture, as the case may be) towards the stable state.
When the levels of annealing are incomplete and, as is often the case,
unequal due to the previous thermomagnetic history, the system will
typically exhibit different levels of correlations and thus different critical currents, as
expected within the LO mechanism. The history dependence of critical
current should thus be a generic occurrence in this regime for comparable
levels of disorder. We emphasize that the proposed explanation of the
various anomalous history dependent magnetization hysteresis data shown
here and elsewhere [9,15,16,19] are in terms of an order-disorder transformation and
disorder-aided {\it supercooling}. This explanation is , prima facie,
independent of the specifics of microscopic considerations
consistent with the ubiquitous nature of the
phenomena under consideration. It is tempting to suggest that the
so-called {\it anomalous} PE in $CeRu_2$ [15,16,18] may also find an explanation within the
scenario described above; whether this is indeed the case remains to be concluded.\\
$^*$ gurazada@apsara.barc.ernet.in or shobo@research.nj. nec.com\\

\begin{center}
{\bf REFERENCES}
\end{center}
\normalsize
\begin{enumerate}
\item  M. Tinkham, Introduction to Superconductivity, second edition,
McGraw-Hill International Editions (1996), U.S.A., Chapter 9.
\item  S. Bhattacharya and M. J. Higgins, Phys. Rev. Lett. 70, 2617 (1993);
Phys. Rev. B 52, 64 (1995); M. J. Higgins and S. Bhattacharya, Physica C 257, 232 (1996)  and references 
therein.
\item  K. Ghosh, S. Ramakrishnan, A. K. Grover, G. I. Menon, G. Chandra, T. V. Chandrasekhar Rao, G. Ravikumar, P. K. Mishra, V. C. Sahni, C. V. Tomy, G. Balakrishnan, D. Mck Paul, S. Bhattacharya, Phys. 
Rev. Lett. 76, 4600 (1996).
\item  L. A. Angurel, F. Amin, M. Polichetti, J. Aarts, P. H. Kes , Phys.
Rev. B 56, 3425 (1997).
\item  C. Tang, X. S. Ling, S. Bhattacharya and P. M. Chaikin, Europhys.
Lett. 35 , 597 (1996).
\item  W. Henderson, E. Y. Andrei, M. J. Higgins and S. Bhattacharya,
Phys. Rev. Lett. 77, 2077 (1996); 80, 381 (1998).
\item  S. S. Banerjee et al, Physica C 308, 25 (1998), Phys. Rev. B 58,
995 (1998); 59, 6043 (1999) and references therein.
\item  T. V. Chandrasekhar Rao et al, Physica C 299, 267 (1998); BARC
(India) preprint 1998.
\item  G. Ravikumar et al, Phys. Rev. B 57, R11069 (1998).
\item  P. L. Gammel, U. Yaron, A. P. Ramirez, D. J. Bishop, A. M. Chang, R.
Ruel, L. N. Pffeifer, E. Bucher, G. D' Anna, D. A. Huse, K. Mortensen, M.
R. Eskildsen and P. H. Kes, Phys. Rev. Lett. 80, 833 (1998).
\item  A. I. Larkin and Y. N. Ovchinnikov, Sov. Phys. JETP 38, 854 (1974); J. Low Temp Phys. 34, 409 (1979); A. I. Larkin, Sov. Phys. JETP 31, 784 (1970).
\item T. Giamarchi and P. Le Doussal, Phys. Rev. Lett. 72, 1530 (1994).
\item T. Giamarchi and P. Le Doussal, Phys. Rev. B 52, 1242 (1995); 55, 6577 (1997).
\item M. J. P. Gingras and D. A. Huse, Phys. Rev. B 53, 15193 (1996). 
\item   S. B. Roy and P. Chaddah, Physica C 273, 120 (1996); J. Phys: Condens. Matter 9, L625 (1997); S. B. Roy, P. Chaddah and S. Chaudhary, {\it ibid} 10, 4885 (1998).
\item  S. Chaudhary, S. B. Roy, P. Chaddah and L. F. Cohen, Proceedings of
the 41st annual DAE Solid State Physics Symposium, Universities Press,
Hyderabad (India), 41, 367 (1998); P. Chaddah and S. B. Roy, Bull. Mater. Sci.,22, 275 (1999).
\item  C. V. Tomy, G. Balakrishnan and D. Mck Paul, Physica C 280, 1
(1997); Phys. Rev. B 56, 8346 (1997) and references therein.
\item M. Tachiki, S. Takahashi, P. Gegenwart, M. Weiden, M. Lang, C. Geibel, F. Steglich, R. Modler, C. 
Paulsen, Y. Onuki, Z. Phys. B 100, 369(1996).
\item S. Kokkaliaris, P. A. J. de Groot, S. N. Gordeev, A. A. Zhukov, R. Gagnon and L. Taillefer, Phys. 
Rev. Lett. 82, 5116 (1999).
\item  S. S. Banerjee {\it et al}, Physica B 237-238, 315 (1997).
\item  G. Ravikumar et al, Physica C  298, 122 (1998) and references therein.
\item  P. Chaddah, K. V. Bhagwat and G. Ravikumar, Physica C 159, 570
(1989).
\item A. K. Grover, in Studies of High Temperature Superconductors, Vol.
14, edited by A. V. Narlikar, Nova Science Publishers, Inc., Comack, NY, USA,
1995, pp.185-244.
\item P. Chaddah, {\it ibid.}, pp 245-273.
\item S. S. Banerjee et al, Appl. Phys. Lett. 74, 126 (1999).
\item Inspection of Fig.3 in Ref.15 (J. Phys : Condens. Matter 10, 4885(1998))
suggests  that  the  minor  loop  initiated  from the reverse leg
indeed overshoots the forward envelope curve even in CeRu$_2$.
\item  M. Steingart, A. G. Putz and E. J. Kramer, J. Appl. Phys. 44,
5580 (1973).
\large
\begin{center}
{\bf FIGURE CAPTIONS}
\end{center}
\normalsize
Fig. 1. Schematic behavior of minor magnetisation curves initiated from
field cooled (I) magnetisation value (M$_{FC}$) and those from forward
(II) and reverse (III) legs of the envelope hysteresis loop within the
framework of critical state model, i.e., assuming J$_c$(H) is uniquely
defined by H. The field interval H$_{II}$, corresponding to a threshold field
change required to change the sign of shielding currents throughout the
sample, is also indicated.

Fig.2. Magnetization hysteresis of a $NbSe_2$ crystal recorded at 6.95K for $H
\parallel c$. The  inset (i) schematically
shows three different paths, viz., the zero field cooled (ZFC), field cooled (FC) and descending fields from above $H_{c2}$. The PE line H$_p$(T) and the upper critical field line H$_{c2}$(T) have been determined from the temperature dependent in-phase ac susceptibility
data, as in Ref.3. Note that in the FC mode, the sample would cross the $H_p(T)$ line 
at different point each time, while reaching a given $(H,T)$ value. Inset (ii) shows  an enlarged view of the $M$-$H$  loop in the PE region, indicating
the onset field $H_{pl}^+$, the peak  field  $H_p$  and
the irreversibility  field  $H_{irr}$.  

Figs.3(a) to 3(c). Minor magnetisation curves in the given $NbSe_2$ crystal at 6.95 K measured along three paths, as schematically sketched in Fig.1. The inset in Fig.3(a) shows
the merger of FC-REV curves initiated from M$_{FC}$(H) (where $H$ $=$ 0.4 kOe and 0.6 kOe) into the reverse envelope curve. Note that the minor curves in Fig.3(b) do not readily overlap with reverse envelope curve, whereas those in Fig.3(c) cut across the forward envelope curve.\\

Fig.4. Field dependence of J$_c$ for $H \parallel c$at 6.95K in $NbSe_2$ for three different histories as indicated.\\
\end{enumerate}

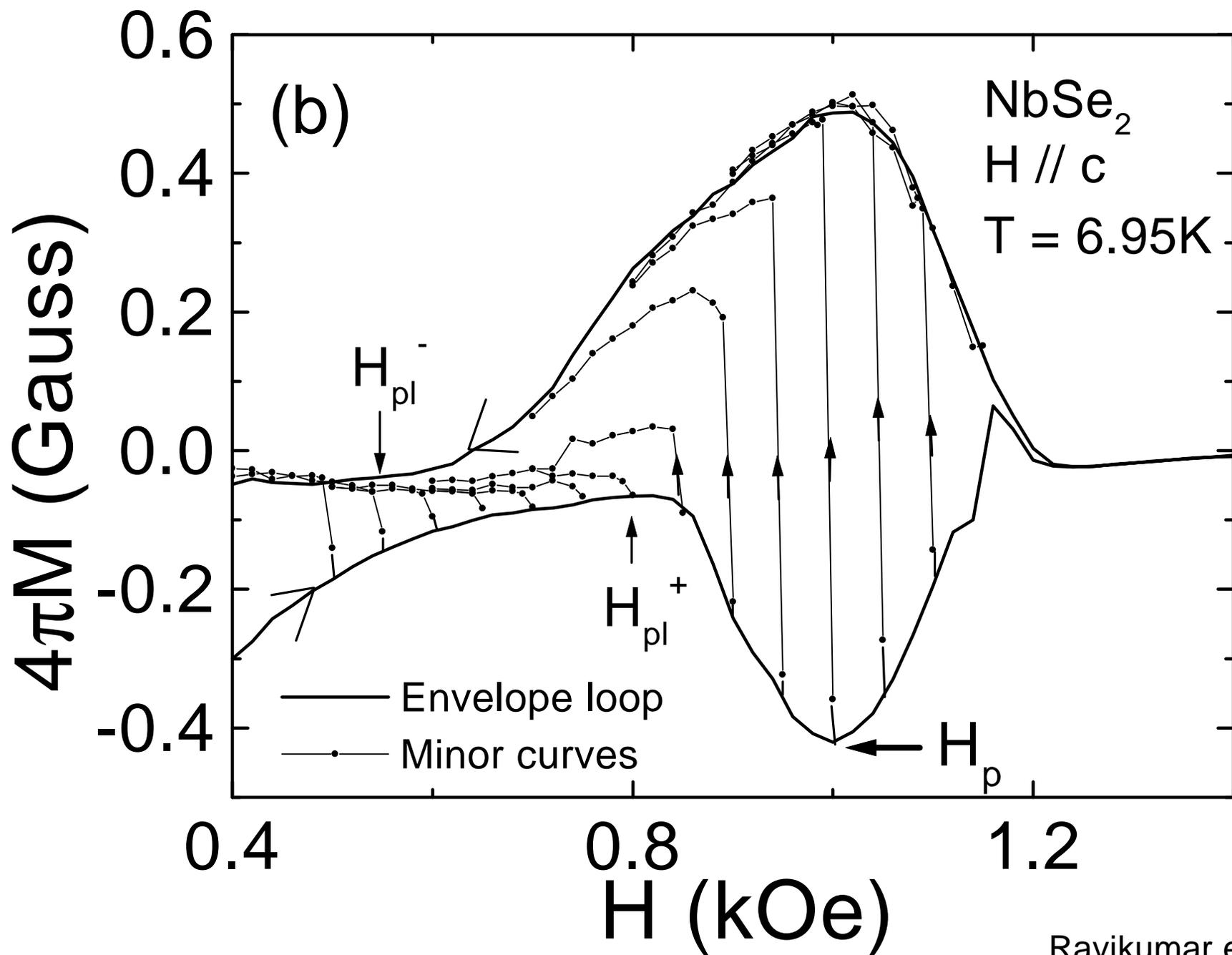
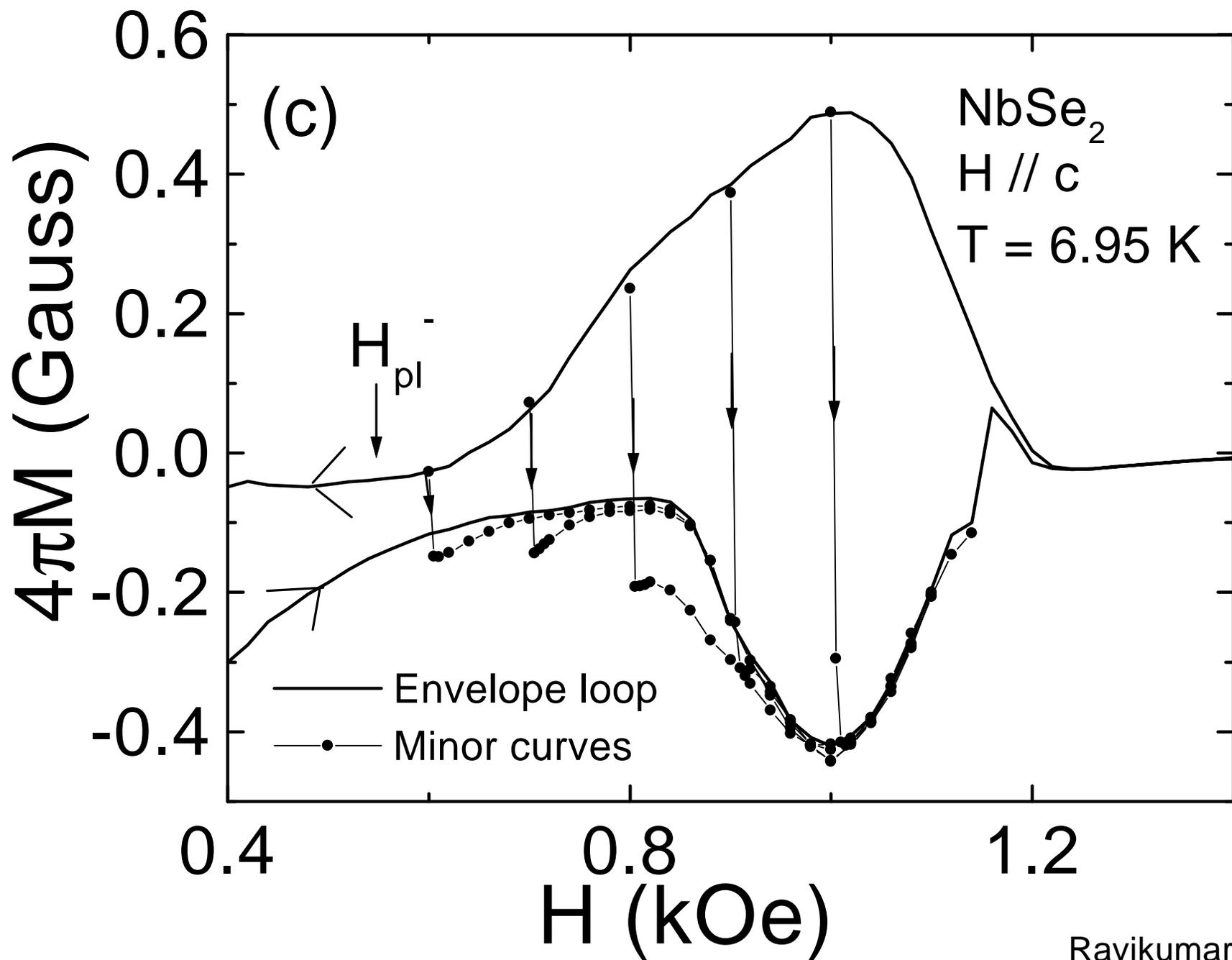
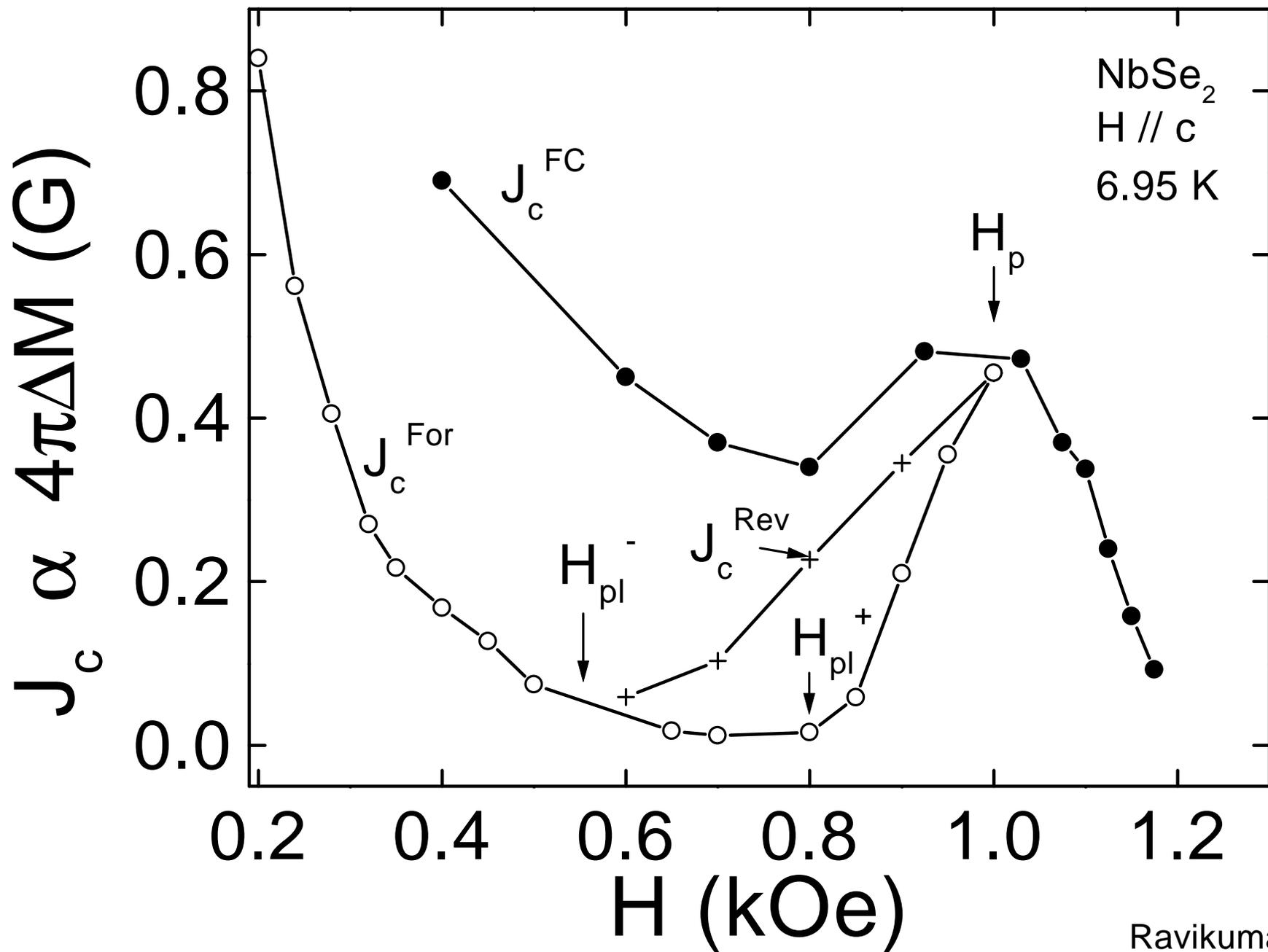
\end{document}